# Quantum Entanglement and Decoherence: Beyond Particle Models. A Farewell to Quantum Mechanics´ Weirdness

O. Tapia
Chemistry Ångström
Uppsala University, Sweden
orlando.tapia@fki.uu,se

Combining abstract to laboratory projected quantum states a general analysis of headline quantum phenomena is presented. Standard representation mode is replaced; instead quantum states sustained by elementary material constituents occupy its place. Renouncing to assign leading roles to language originated in classical physics when describing genuine quantum processes, together with sustainment concept most, if not all weirdness associated to Quantum Mechanics vanishes.

**Introduction**

Quantum entanglement and coherence hang around as physical phenomena holding keys to apprehend the nature of a quantum world. Not only chemical dynamics and changes of chemical bonds patterns, but also coherent energy transfers in biological systems are cases directly concerning such quantum states.

Using a quantum scheme where Hilbert space mathematics is untouched and modifying foundational quantum tenets that now differ from representational ones, quantum physical processes are formulated in terms going beyond semi-classic schemes debasing features such as potential energy surfaces, particle models and representation mode.[1-9] Representation by particles is a characteristic of classical physics systems; such mode is banned from the present framework and replaced by the concept of quantum states *supported* by a given materiality[1,5] Therefore, an entangled state does not represent a property; properties are characteristics of particles in classic and semi-classic physics.

With focus shifted away from the material sustaining q-states (quantum states), entangled states turn out to play central roles in rationalizing quantum processes. Such scheme might help constructing theoretically consistent views to bridging in near future abstract states to computational quantum chemical schemes. Thus, multipartite bases obtain that show up to be useful to describe quantum states of complex systems.[1-3] The nature of the quantum state stands out as key difference with previous approaches.

At laboratory level, basic material elements (e.g. fixed numbers of electrons and/or nuclei, or even quantum dots) *sustain* the q-states; yet these ones do not represent in a classical physics sense such objects. Q-states subsume information to be experimentally gleaned and/or modulated overcoming the idea of particles occupying (base) states;[5,6] actually, all possibilities accessible to a given system must be reckoned in the scheme via multipartite basis sets.[1-3]

That time and space are central to the scheme is evident. Quantum entanglement raises important queries on locality that imposes an attentive and fresh analysis of configuration space framework, e.g.[2,3] On the one hand, abstract quantum states should handle all response-possibilities available to physical systems. On the other hand, probabilities emerge from laboratory demolition measurements, namely *events* at the fence of two worlds.[5,6] These events in the present framework would convey information richer than just counting sequences.[2,3,7] Thus, a more functional view will replace the standard representation mode of quantum mechanics; and, as a consequence, weirdness is left behind.

Classical physics elements are identified by mappings between abstract and laboratory space elements, basically introducing inertial frames (I-frames) from special relativity theory (SRT) and transformation groups invariances; these elements (e.g. quantum numbers) are hence merged in quantum frameworks with algebraic (graded) models and suitable algorithms helping calculate base states sustained by electro-nuclear (EN) elements;[6] the *information* so gleaned is re-injected as quantum labels.[3,5-8]

**Entanglement and Probing**

Entanglement [10] catches up a quantum foundational notion that concerns quantum states as such and not directly elementary materiality (particles) sustaining them.

Consider the states $|\pm\rangle_{12}$ given as coherent linear superpositions for a bipartite system:
$$|\pm\rangle_{12} = (1/\sqrt{2})[\langle\mathbf{x}_1|f_{\mathbf{k}'1}\rangle\otimes\langle\mathbf{x}_2|f_{\mathbf{k}'2}\rangle \pm \langle\mathbf{x}_1|f_{\mathbf{k}'2}\rangle\otimes\langle\mathbf{x}_2|f_{\mathbf{k}'1}\rangle] \quad (1)$$

$|\pm\rangle_{12}$-equation corresponds to two non-separable (and orthogonal) modes, $|+\rangle_{12}$ and $|-\rangle_{12}$ where separate elements are not

directly accessible if (1) stands for a robust feature. Configuration space labels {$x_i$ (i= 1,2)} referred to an I-frame are put in correspondence (map) to abstract point coordinates; these labels chart to the number of classical degrees of freedom (DOF) not to locations.[7,8] The I-frame origin is label from another I-frame so that relative I-frame motion in laboratory space may be either classically described or incorporated in box-like quantization.[4-7]

To complete the basis set over 3m-dimension DOF, $<x_3,…,x_m|f_{k'3…k'm}>$ stands for base states supported on 3m-6 classical DOF; quantum DOF are given as quantum numbers ($k'_1…k'_m$). Q-states $|\pm>$ basis takes on the form of row vector:
(…$<x_1|f_{k1}>\otimes<x_2|f_{k2}>\otimes<x_3,…,x_m|f_{k'3…k'm}>$…
…$<x_1|f_{k2}>\otimes<x_2|f_{k1}>\otimes<x_3,…,x_m|f_{k'3…k'm}>$…)

The base set naturally includes quantum labels permutations; retaining two generic components only, entangled state (1) takes on the form:

(…$<x_1|f_{k1}>\otimes<x_2|f_{k2}>\otimes<x_3,…,x_m|f_{k'3…k'm}>$
$<x_1|f_{k2}>\otimes<x_2|f_{k1}>\otimes<x_3,…,x_m|f_{k'3…k'm}>$…)•
(…$1/\sqrt{2}$   $\pm 1/\sqrt{2}$…)$^T$ → $|\pm>_{12}$        (2)

The scalar product reads then as (1) though multiplied by $<x_3,…,x_m|f_{k'3…k'm}>$. It is in the transposed vector with amplitudes (…$1/\sqrt{2}$ $\pm 1/\sqrt{2}$…)$^T$ that actually stands for the quantum state, the base set is always invariant. There is a caveat discussed later on.

Let a probing device on the one hand induce a transition relating states $|+>$ to $|->$ at a given space location: a scattering source, or detector/ register would do the job; on the other hand, the partite elements in (1) are implicitly referred to particular I-frames each. Thus, real measuring devices would map states from abstract Hilbert to laboratory space (lab-space); take a resultant quantum state after the physical interaction operates to be either $(1/\sqrt{2})$ $(|+>+|->)$ or $(1/\sqrt{2})(|+>-|->)$. This case expressed as two possible yet exclusive laboratory-related results:

Transition $\hat{T}^+|+>$: $(1/\sqrt{2})(|+>+|->)$ →

$<x_1|f_{k1}>\otimes<x_2|f_{k2}>\otimes<x_3,…,x_m|f_{k3…km}>$
(3a)

Transition $\hat{T}^-|+>$: $(1/\sqrt{2})(|+>-|->)$ →

$<x_1|f_{k2}>\otimes<x_2|f_{k1}>\otimes<x_3,…,x_m|f_{k3…km}>$
(3b)

Measurement via actual probing senses states related to either (3a) or (3b). Three I-frames result in this picture for each element of the direct product; we can now *locate* two devices in lab-space. Detectors located, say at $R_1$, $R_2$, respectively. By sensing a response rooted at state $<x_1|f_{k1}>$ from (3a) at say $R_1$ a simultaneous response ought to be collected from base state $<x_2|f_{k2}>$ at conveniently chosen $R_2$ (see below)
If these possibilities concretize, probing results pinched at $R_1$ and $R_2$ must be strongly correlated. More interestingly, successive measurements will show either possible responses from $<x_1|f_{k1}>$ or $<x_1|f_{k2}>$ with a limiting square modulus amplitude value of ½ after large trial numbers.

These are possibilities accessible to the system interacting with a real probing device. The physical effect resulting from probing, if successful, will *move amplitudes* from the entangled base component $|\pm>_{12}$ to a non-entangled one i.e. (3a) or (3b). The entangled state is hence "destroyed" while materiality remains invariant. A real interaction imparts then I-frames as it were to the elements of the direct product.
*In this case probing (measuring) changes number of partite elements by adding two. This is commensurate with a dissociation process when referred to coherent state (2).*

Thus an important difference between (1) and (2) becomes apparent. Examine the situation in more detail to improve apprehending this laboratory quantum measurement.
Let partite base state (2) serve as reference I-frame in uniform state of motion (conserved quantity); in lab-space associate I-frames to each partite element $<x_1|f_{k'1}>$ and $<x_2|f_{k'2}>$ to mimic actual response; select the case where in k-space I-frames displace in opposite directions: $\hat{k}_1 = -\hat{k}_2$, independently from internal quantum numbers: the pairs $<k_1|f_{k1}>\otimes<k_2|f_{k2}>$ and $<k_1|f_{k2}>\otimes<k_2|f_{k1}>$ are obviously equivalent.
Moreover, taking these elements with respect to $<x_3,…,x_m|f_{k'3…k'm}>$, i.e. the anchor (m-2)-partite state, their locations are opposite (antipodes, the 3 I-frame origins in a line); this results from conservation of linear and angular momentum; I-frames axis orientation is arbitrary if the outgoing state is a S-state (spherically symmetric). All these statements refer to *accessible possibilities*. Not to particles.

Take now any two correlated points in space locating detectors fulfilling constraints just stated; consider for instance, a response rooted at $<x_1|f_{k'1}>$ and the alternative one necessarily at $<x_2|f_{k'2}>$. These responses would click simultaneously as they belong to an entangled state; similarly for case $<x_1|f_{k'2}>$ and $<x_2|f_{k'1}>$. In a sense, linear and angular momentum conservation elicited by physical processes will tie up materiality *response* (to a location *so to speak*), the reason is simple: because materiality sustains the quantum state at the end you cannot get one without the other.

What happens if only one detector is used?
We are now closer to Einstein-Podolski-Rosen (EPR) original paradox,[19] although with a caveat. For, if one assembles only one sensor, (an object one commands from outside) a local probing of the entangled state with effective detections, namely, detecting at de particular location a response rooted say at $<x_1|f_{k'1}>$, implies that at the antipode there *must* be a "virtual" response from state $<x_2|f_{k'2}>$ *whether you measure it or not*. If you, as EPR did, understand all these in terms of (independent) particles there seems to be a "spooky" action at a distance produced by detection at one point in laboratory. And all adjectives on the weirdness of quantum mechanics "naturally" appear in describing events.

But a problem here is that in this *kind of talking* one mixes possibilities with actualities: this is a misunderstanding or a misinterpretation at best. In fact, one ought to analyze the possibilities first and thereafter set up the probing systems (in real space) and accordingly select subsets *compatible under experimental constraints*. And remind that it is to quantum states that probing addresses, not to objects (molecules, particles or clicks).

Let come back to the generic quantum state to be probed. Formally, all three partite elements firstly share the I-frame as they appear as terms of a coherent superposition, namely (2).
The entangled state $|\pm>_{12}$ belong to a S-state means that is associated to an infinite number of possibilities corresponding to radially "propagating" I-frames states. This is a particular possibility-space.
Therefore, the states that can be sensed at the tips: $\hat{k}_1 \xleftrightarrow{=} -\hat{k}_2$ actually are the same abstract state eq.(1) or (2). Spherical symmetry associated to a possibility-space means that at any point on a sphere drawn with the help of a source and location of a putative detector there is a "copy" of state (1). This situation characterizes *abstract spaces*.
Actual probing would effect on quantum state (2) a change that we express as (3a) or (3b). So that, getting response at a location we are informed that it is simultaneously manifested at an antipode location.

Viewed from a classical physics perspective the situation just described is a conundrum. *Yet, from a quantum physical viewpoint one probes one and the same state, namely, an entangled state*.

Further precisions:
(i) Once entanglement sets up, quantum state (1) is the same at any configuration space point as well as (parametric) time;
ii) The quantum state under laboratory conditions is sustained by a given materiality (e.g. number of electrons and nuclei);
iii) Even if no materiality is present, the state remains to the extent this is an *abstract* state (a possible state) also, *no actual response for so long materiality is absent*. This latter point may seem strange. But, quantum physics is about possibilities and one of them may be absence in the volume space of the sustaining material (vacuum). This is no more complicated than saying: whatever is done, no real physical response from the system rooted at the given quantum base state is expected if no materiality shows up there to actually sustain it; this is what *presence* refers to. *This is a characteristic of laboratory physical states under probing*.

Thus, materiality's presence is a key to physical responsiveness.[5,6] The other way round, response put in evidence particular materiality (with caveats related to noise).

**Entanglement View from Abstract Quantum Frameworks**

As the preceding discussion makes clear, (1) retains a particle-like flavor implied by the dominating view of configuration space. Deeply rooted in our teaching of Quantum Mechanics lies the particle representation idea. For this reason, there is need to bypass a representational mode. This obtains by using the extended abstract basis set as described in our papers:[1-3,6-8]

$|\pm> \rightarrow$ (2')
$(0_{1\text{-Partite}} \ldots 0_{2\text{-Partite}} \ldots 1/\sqrt{2} \ldots \pm 1/\sqrt{2} \ldots 0_{m\text{-Partite}})^T$

The basis set includes now all partitioning possible that a finite number of elementary material constituents may sustain.[1-3] The state (2') selects a subset from the base vector that is given as:

(|1-Partite>…|2-Partite>…<$x_1|f_{k'1}$>⊗ <$x_2|f_{k'2}$> ⊗ <$x_3,…,x_m|f_{k'3…k'm}$>…<$x_1|f_{k'2}$> ⊗ <$x_2|f_{k'1}$>⊗<$x_3,…,x_m|f_{k'3…k'm}$>…|m-Partite>)  (4)

As indicated in state vector (2') the amplitudes affecting 1-,2-,4-…m-Partite bases are put equal to zero to follow up our model states now in an explicit infinite dimensional Hilbert space. States $|\pm>$ in (2') are clearly not separable with respect to $|\pm>_{12}$ and <$x_3,…,x_m|f_{k'3…k'm}$>; it is a coherent linear superposition referred to only one I-frame.

If we use $\hat{T}^+$ to present a probe effect as a change of state vector, one formally gets:

Transition $\hat{T}^+|+>$:
$\hat{T}^+(0_{1-Part}…0_{2-Part}…1/2…+1/2…0_{m-Part})$ =
$(0_{1-Partite}…0_{2-Partite}…1/2+1/2…1/2-1/2…0_{m-Partite})$ =
$(0_{1-Partite}…0_{2-Partite}…1…0…0_{m-Partite})$    (5a)

Transition $\hat{T}^-|+>$:
$(0_{1-Partite}…0_{2-Partite}…1/2-1/2…+1/2+1/2…0_{m-Partite})$ =

$(0_{1-Partite}…0_{2-Partite}…0…1…0_{m-Partite})$    (5b)

One remains in Hilbert space; it is apparent that no "collapse" is possible. The basis stay invariant and only the components of the state vector changes.
In order to move towards lab-space, external actions are required (see below).

**Decoherence**
A new phenomenon becomes apparent at this point: lab-space de-coherence that is originated by real space interactions. Surroundings effects so to speak.
Projecting the abstract state correspond to making explicit the scalar product so that one can get projections (3a') and (3b'):
$\hat{T}^+|+>$ → (|1-Partite>…|2-Partite>…<$x_1|f_{k'1}$>⊗ <$x_2|f_{k'2}$> ⊗<$x_3,…,x_m|f_{k'3…k'm}$> … <$x_1|f_{k'2}$>⊗ <$x_2|f_{k'1}$>⊗ <$x_3,…,x_m|f_{k'3…k'm}$> … |m-Partite>)•
$(0_{1-Partite}…0_{2-Partite}…1…0…0_{m-Partite})^T$
The projection leads to:
<$x_1|f_{k'1}$>⊗<$x_2|f_{k'2}$>⊗ <$x_3,…,x_m|f_{k'3…k'm}$>     (3a')
Or, by using the transposed vector (5b):

<$x_1|f_{k'2}$>⊗<$x_2|f_{k'1}$>⊗ <$x_3,…,x_m|f_{k'3…k'm}$>    (3b')

Note that (3a') and (3b') each one refers to only one I-frame, they are therefore not independent elements and consequently they cannot be sensed at different locations. Yet they share the same form as (3a) and (3b).

Abstract quantum theory does not describe materiality whereabouts in real space as if they were independent elements; thence no trajectories.[7,8]

So long entangled states evolve in *abstract* space all possibilities are accessible and can be calculated; some of them might be meaningless (to us) but could be used to help apprehend aspects of phenomena.
But, for the present case, entanglement means that *once materiality is detected*, the result (namely a detection or click) implies that both I-frames are unraveled by $T^\pm$ successful interactions.
Henceforward, a $T^\pm$ transition connects spaces with different number of I-frames; among other cases, it might be used to activate clocking; it requires a quantum energy exchange between the detector and physical quantum states characterizing probing.[9] Thereafter parametric time should leave the place and be replaced by laboratory time.

Interestingly, the situation above precisely characterizes decoherence. Here lies another of quantum physics conundrum if a classical physics viewpoint is used to describe it.

The entangled quantum state has a special trait then: either it expresses as possibility or does it as presence *concomitantly* with probing elicited by a particular event (or a cascade).
If only one detector is present and if it were activated at a given laboratory time, the nature of the interaction, implying presence of the entangled state, would be triggered if and only if second materiality component is present too, simultaneously yet not measured; otherwise, detection could not be actualized. This *information* comes from the entangled state. It belongs to a world that is included in a quantum world.
The passage from one to three I-frames is a process happening in laboratory space. This implies that probe and probed systems must be included as well.[9] In this sense standard Quantum Mechanics is incomplete.

Furthermore, as already discussed, relevant materiality must be present at laboratory level to ensure conservation laws once for instance a T$^+$ probing successfully takes place. The one way (and possibly the only one in absence of noise) to get a response at detector is to realize an interaction with *direct intervention of materiality sustaining the quantum state*.

At this point, and only at this one, I-frames over partite states in a manner of speaking become apparent (to us); *this would be a result of (real) interaction*;[5,6] it cannot happen in abstract Hilbert space. The way it happens at the Fence is beyond reach.[7,8]

Real space interactions cost money as signaled for instance by detections of Higgs states. In other words, real events bring forth local space time; and only in this sense events can be given a dynamic property.

Note that, before probing, there is no requirement for signal "transmission" between terms of the entanglement (1), no spooky action at a distance; actually there is no distance; direct products elements keep topologic relation among them. Materiality implied by the entangled state must be "at the right place" at the "right time," the procedure amounts to prepare the quantum state for the unperturbed (non-probed) partite state. This is the meaning of concept: *quantum- state- sustained- by- a- given- materiality*.

If one looks at the semi-classic case discussed by EPR a simple result follows: for them, both "particles" are present but *knowledge* of their state is missing. Detecting the state for one of them the second appears to be *enforced* at a distance in real space that may be beyond what it is allowed by SRT, a sort of superluminal signal. From the viewpoint used here, there is no such enforcement; *the state was already there*. Superluminality just fade away and with it goes another weirdness dashed.

The quantum formalism imposes action to be simultaneous; yet there is no real space involved and no real time spent. Such is the nature of quantum entanglement. *It provides the experimenter with a resource*. A resource found at the grounds that sustain the idea of quantum computers as well as teleportation of quantum states (not particles I'm afraid).

Another problem with classical analyses is that it separates the entanglement terms and treats them independently of each other occupying locations in real space. Moreover, quantum physics does not address particle's motion; *I-frame motion* is the classical link. Time dependent amplitudes stand for q-dynamics yet not particle dynamics.

From the preceding descriptions one can conclude that: tenets of classical physics are not even wrong in quantum-entangled circumstances; they are irrelevant at best. Thus:

*Quantum entanglement cannot be simulated in classical physics terms*.

This conclusion closes a first view of quantum entanglement from the perspective developed here and elsewhere.[1,7,8] It agrees with Bell's 1964 theorem, stating that the predictions of quantum theory cannot be accounted for by any local theory; this theorem represents one of the most profound developments in the foundations of physics.[11]

The agreement of the present approach with this important theorem is rewarding and suggestive. Note, our approach concerns *not only* event counting but also and foremost amplitude's sensing. In other words, measuring wave functions would be a requirement imposed when thinking about and not only (irreversible) events detection. Thus, putting together most (if not all) possible events one gets a "portrait" of a quantum state.

Any statistical interpretation seems to be inadequate.

**Discussion**

The paper addresses some basic problems in quantum physics and quantum mechanics in particular. In opinion of Steven Weinberg there is no entirely satisfactory interpretation of quantum mechanics.[12] Besides, two-photon interferometry in recent Hobson's [13] work corroborates Weinberg's idea; he discusses collapse of quantum state in a manner differing from the standard view. Yet grounding elements used, such as particles and representational mode, remain in widespread use. This is surprising since the question rose by Laloë [14] asking whether one really understands quantum mechanics still remains unanswered.

The present work and those reported in [1,5,6] credit the idea that we are rather far from understanding quantum mechanics if we insist using classical physics tenets; the oxymoronic character of wave-particle duality is difficult to hide. The ideas presented here radically differ from dominating archetypes. The discussion of entanglement presented above illustrates this point. Entanglement concerns q-states in the first place, not the elementary material system sustaining it. Consequently an interpretation of quantum mechanics, if that one is to refer to the materiality sustaining quantum states, is expendable.

Yet, *event* sequences as such, e.g., those recorded under laboratory premises, can be submitted to probabilistic analyses; they share an "objectiveness" character (I-frame sustained). In this case the weight is put on recorded material: the spot (click). In this context it is reasonable that Bayesian analyses [15] turn out to be useful since the enquiring direction is from outside (laboratory) to the inside (quantum). However, the supporting role of materiality with respect to quantum states, that is fundamental for the present view, fades away in a probabilistic (counting) context. And, consequently quantum mechanics' weirdness's would show up again.

To bypass such contradictory way of speaking we ought to accept that Classical physics and Quantum physics languages are, strictly speaking, irreconcilable. Yet they can still be used in an harmonizing manner with the classical one playing a subsidiary almost didactic role; the introduction of inertial frames with partite base sets is one example.[1-4]

Entanglement is not a property of materiality defining the physical system. It is a quantum state that expresses a peculiar set of responses once projection from abstract to laboratory domains is done. It is definitely not an object and consequently it is not a property in the classical sense. As noted above, it is a resource.

Speak of entanglement and disentanglement in abstract space is *formally* possible as the set of amplitudes changes in a well-defined form: compare (2') and (5a),(5b).
To produce an entangled state from two space-separated and uncorrelated partite states requires a third partite element, e.g. a beam splitter. One with double input can produce partially entangled output quantum states. A beautiful example is found in the work by Olmschenk et al.[16]
In the same vein, a covalent chemical bond expresses an entangled state.[1-3] Two partite states corresponding to independent fragments require of a "third body" to put amplitudes at the state standing for chemical bond; in this sense we do not "break" neither "knit" bonds. Take a sample of hydrogen atoms.[4] At very low temperature they form Bose-Einstein condensates. To put amplitudes at the one partite state corresponding to an hydrogen molecule the container surface can catalyze the entanglement/disentanglement process in real space.[4]

Examples above illustrate entanglement as a resource to actually apprehend chemical and physical process at abstract lab-space before decoherence takes over. A proper handling of q-states prevents use of the elements of entangled states as if they were independent. Probability concept is not suitable before decoherence has set up.

**Acknowledgments**
The author is much indebted to Uppsala University and Valencia University for all resources put at his disposal.